\documentclass[10pt,twocolumn,prl,aps,superscriptaddres,floatfix]{revtex4-1}

\usepackage{amssymb}
\usepackage{amsmath}
\usepackage{bm}

\usepackage[linkcolor = blue, citecolor = blue, urlcolor = blue, colorlinks = true]{hyperref}

\usepackage{graphicx}
\usepackage{wrapfig}
\usepackage[dvipsnames]{xcolor}

\usepackage{comment}
\usepackage[utf8]{inputenc}
\usepackage{lipsum}

\begin{document}

\title{Structuring Colloidal Gels via Micro-Bubble Oscillations}

\author{K.W. Torre}
 \email{k.w.torre@uu.nl}
\affiliation{
 Institute for Theoretical Physics, Center for Extreme Matter and Emergent Phenomena, Utrecht University, Princetonplein 5, 3584 CC Utrecht, The Netherlands 
}
\author{J. de Graaf}
\affiliation{
 Institute for Theoretical Physics, Center for Extreme Matter and Emergent Phenomena, Utrecht University, Princetonplein 5, 3584 CC Utrecht, The Netherlands 
}

\date{\today}


\begin{abstract}
Locally (re)structuring colloidal gels --- micron-sized particles forming a connected network with arrested dynamics --- enables precise tuning of the micromechanical and -rheological properties of the system. A recent experimental study [B. Saint-Michel, G. Petekidis, and V. Garbin, Soft Matter \textbf{18}, 2092 (2022)] showed that rapid restructuring can occur by acoustically modulating an embedded microbubble. Here, we perform Brownian dynamics simulations to understand the mechanical effect of an oscillating microbubble on the structure of the embedding colloidal gel. Our simulations reveal a hexagonal-close-packed restructuring in a range that is comparable to the amplitude of the oscillations. However, we were unable to reproduce the unexpectedly long-ranged modification of the gel structure --- dozens of amplitudes --- observed in experiment. This suggests including long-ranged effects, such as fluid flow, should be considered in future work.
\end{abstract}

\maketitle


\section{\label{sec:intro}Introduction}

A colloidal suspension can gel~\cite{lekkerkerker1992poon}, when short-ranged attractions --- typically induced by the presence of polymers~\cite{asakura1958interaction} --- are much larger than the thermal energy $k_{\mathrm{B}} T$; here, $k_{\mathrm{B}}$ is the Boltzmann constant and $T$ the temperature. These attractions arrest the system's natural tendency to (spinodally) phase separate, leading to the formation of an open, space-spanning network structure that is intrinsically out of equilibrium~\cite{zaccarelli2007colloidal, Royall2021}. The network structure can for a finite (often long) time support the gel's buoyant weight against gravity~\cite{zaccarelliHarich2016, starrs2002collapse}. Stability at low volume fraction, has led to the widespread use of particle gels in industrial, medical, and academic settings,~\textit{e.g.}, care products, printing
inks, foodstuffs, crop protection, and pharmaceutical suspension formulations~\cite{larson1999structure, eryt, food-soft-materials, crop_protection}. This has led to scientific interest in the properties of colloidal gels, and such systems have been studied using experimental~\cite{carpineti1992spinodal, verhaegh1999transient, cipelletti2000universal, starrs2002collapse, poon2002physics, shah2003microstructure, manley2005time, krishna2012probing, bartlett2012sudden, zaccarelliHarich2016, razali2017effects, tsurusawa2019direct}, computational~\cite{foffi2002evidence, del2003unifying, puertas2004dynamical, zaccarelli2009colloidal, furukawa2010key, vargaswan2015, varga2018normal, varga2018modelling, padmanabhan2018gravitational, swan-furst2019, gelhydroJoost2019, majji2020hydrodynamic}, and theoretical~\cite{buscall1987consolidation, lekkerkerker1992poon, allain1995aggregation, allain2001systematic, starrs2002collapse, bergenholtz2003gelation, chen2004microscopic, weitz2005gravitational} methods.

Gels coarsen over time, as the system relaxes toward equilibrium, and their bulk properties can strongly depend on the preparation history~\cite{gelhistory}, including oscillatory-shear \cite{tuning_oscillatory_shear1, tuning_oscillatory_shear2}, and steady-shear protocols~\cite{tuning_steady_shear}. That is, the preparation can leave a clear signature in the microstructure of the gel~\cite{tuning_oscillatory_shear1}, which expresses itself in the mechanical response of the material~\cite{sudreau2022shear}. Modifying a gel's properties \textit{via} external means has mostly focused on the bulk response. However, for many processes, it can be favorable to apply these modifications locally both for colloidal~\cite{localtuning} and other types~\cite{polymer_gel_expansion} of gel.

Recently, Saint-Michel~\textit{et al.} showed that the dynamics of a deformable inclusion, taking the form of a (micro)bubble, can be used to locally tune a gel's microstructure~\cite{Garbin-gel}. In these experiments, ultrasound is used to cause the bubble to contract and expand, leading to an extensional driving of the surrounding gel. The study revealed a non-trivial rearrangement of the colloidal network into a crystalline structure. The most interesting feature being the long range --- comparable to the bubble radius --- over which the rearrangements took place, when only small oscillations ($\sim 1\%$ of the bubble diameter) are employed. The exact physical mechanism behind this long-range rearrangement remains unclear. Locally perturbing the system using ultrasound and air inclusions can also be useful to probe the rheological response at the scale of the microstructure~\cite{poulichet2015ultrafast, huerre2018dynamic, microrheology_bubble}.

In this work, we use computer simulations to investigate bubble-oscillation based local reordering of colloidal gels. Our model is based on an effective, Asakura-Oosawa-like description of depletion interactions between the colloids, following an earlier analysis of gelation~\cite{gelhydroJoost2019}. The microbubble is described using a bead-spring model subjected to an external (radial) forcing that models pressure changes, due to the ultrasound. We take into account only the mechanical interactions in our model,~\textit{i.e.}, we ignore hydrodynamic interactions between the colloids and porous-medium flow.

For experimentally relevant colloid volume fractions, we vary the colloid-bubble size ratio, the frequency, and amplitude of the oscillations. This allowed us to construct a state diagram that highlights the effect of the oscillations. We find that crystalline reordering into a hexagonal closed-packed state around the bubble is possible, whenever multiple layers of colloids are compressed by the extensional driving of the bubble, and the frequency of the oscillations is large enough to avoid extraction of colloids from the gel network. Turning to the range of the rearrangements, our analysis reveals that this is roughly twice the amplitude of the oscillation. This suggests that there is a missing ingredient to understanding the experiment. However, the present study lays a solid foundation for future work in this direction.

The rest of this paper is organized as follows. We first introduce our numerical method. Next, we cover how we analyse our results, before we show the phase diagram. This is followed by a discussion of the relevant time scales and an outlook on follow-up studies. 

\section{\label{sec:meth}Numerical Method}

We want to study the influence of an oscillating microbubble on the microstructure of a colloidal gel. We do so by performing Brownian dynamics simulations. These take into account the friction between colloids and solvent at a one-body level,~\textit{i.e.}, the Stokes drag, ignoring any two- or many-body interactions. We also ignore any flows in the gel network that might be caused by motion of the gas-liquid interface.

The overdamped equations of motion for a single colloid in our system can be written as:
\begin{align} \label{eom}
    \gamma \frac{ \partial \boldsymbol{r}_{i}}{ \partial t} = \boldsymbol{F}^{p}_{i} + \sqrt{2 k_{\mathrm{B}} T \gamma } \boldsymbol{\xi}_{i},
\end{align}
with $\boldsymbol{r}_{i}$ the $i$th colloid's position. The prefactor $\gamma = 3 \pi \eta \sigma$ specifies the fluid friction experienced by a single colloid, assuming here the Stokes form for a sphere with $\eta$ the viscosity. The forces acting on the $i$th colloid, which derive from pair interactions with neighboring colloids \textit{via} the potentials specified below, are given by  $\boldsymbol{F}^{p}_{i}$. The term $\boldsymbol{\xi}_{i}$ accounts for thermal fluctuations, which are independent and have a white-noise spectrum. That is, we ensure a zero mean $\langle \boldsymbol{\xi}_{i} (t) \rangle = \boldsymbol{0}$ --- the angled brackets indicate a time average --- and  $\langle \boldsymbol{\xi}_{i} (t) \otimes \boldsymbol{\xi}_{j} (t') \rangle = \delta_{ij} \delta( t - t') \mathbb{I}_{3}$. Here, $\otimes$ indicates the tensor product, $\delta_{ij}$ represents the Kronecker delta, $\delta(t - t')$ the Dirac delta, and $\mathbb{I}_{3}$ is the three-dimensional identity matrix.

\begin{figure}[!htb]
 \centering
 \includegraphics[width=28.5mm]{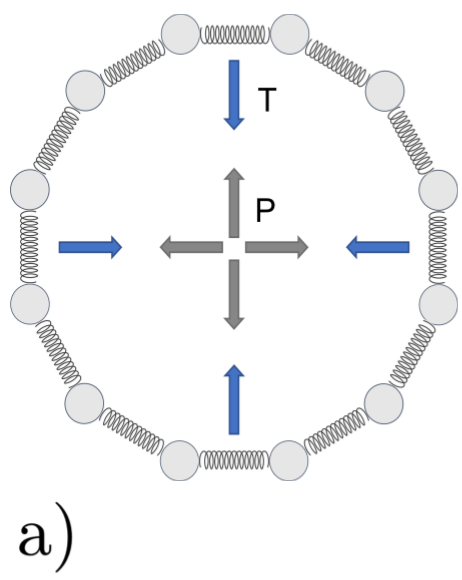}
 \includegraphics[width=49.5mm]{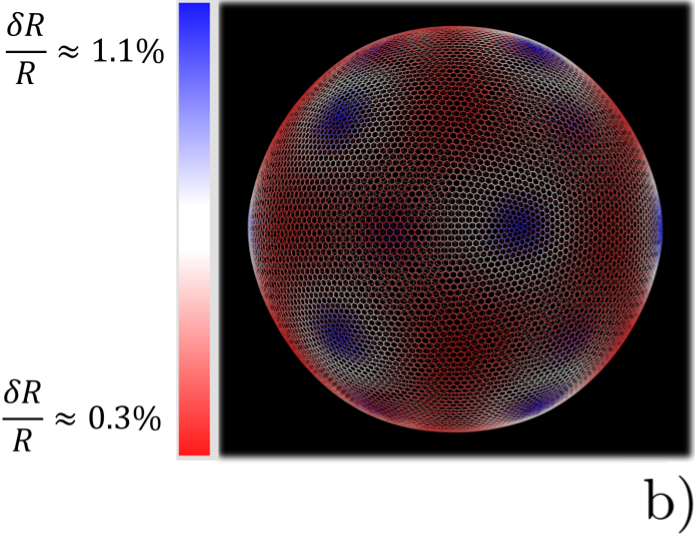}
 \caption{\label{fig:schematic}Model for the oscillating microbubble. (a) Two-dimensional sketch of the bead-spring model. Radially outward `pressure' forces (gray) counterbalance inward contraction from tangential springs (blue) representing `surface tension'. (b) Snapshot of one of our geodesic spheres at rest. The small non-uniformity of the radius induced by point defects, expressed as $\delta R$, is highlighted using the colors in the legend. Blue areas (centered around the 12 pentagonal defects) have a slightly larger radius than the mean value $\langle R \rangle$, while red areas have a smaller radius.}
\end{figure}

We model the microbubble as a collection of points that define a geodesic polyhedron. The facets spanned by the vertices represent the bubble surface. We emulate the internal pressure by adding a constant outward-pointing force acting on each vertex. Surface tension is modelled by connecting neighboring vertices \textit{via} harmonic springs. The spring constant and \textit{equilibrium} pressure are tuned in such a way that the bubble has a mean radius of $\langle R \rangle$ at rest. The energy scale associated with the spring constant $k_s = 10^{7} k_{\mathrm{B}}T \sigma^{-2}$, where $\sigma$ is the colloid diameter, and we used an equilibrium pressure $p_{0}=\Omega k_s / 4\pi \langle R \rangle$, with $\Omega \in [0.22,3.14] 10^{-3}$ a dimensionfree coefficient given by the bubble tessellation. Our choices ensured that when the bubble oscillates, it forces the gel out of the way sufficiently vigorously not to cause distortions in its (nearly) spherical shape, in line with the experimental observations. The resulting model bubble is represented in Fig.~\ref{fig:schematic}.

We had to account for topological constraints imposed by working with a spherical surface, namely that it cannot be tessellated with hexagonal tiles only, 12 pentagonal defects must be present~\cite{goldberg_polyhedra}. These defects introduce distortions away from perfectly spherical in our bubble surface, as can be appreciated from the coloring in Fig.~\ref{fig:schematic}. Red denotes a depression in the bubble surface with respect to its mean radius, whilst blue indicates an increase of the radius. The effect is exaggerated in our representation, as the deviations are typically less than $1\%$. In constructing our geodesic sphere, we have ensured that the defects are located on the vertices of an icosahedron. This localization is convenient, as it allows us to take slices between the defects, where the change in curvature is minimal. In total there are six such slices possible, which proved sufficient to perform a quantitative analysis of the gel, which we will return to shortly.

We modeled the gel according to the methods detailed in Ref.~\cite{gelhydroJoost2019}. In brief, we simulate only the colloids and account for the presence of the polymers that cause depletion attraction via a generalized ``high-exponent'' Lennard-Jones potential
\begin{align}
    V_{\mathrm{LJ}}^{\mathrm{he}} &= \epsilon \left [ \left ( \frac{\sigma}{r} \right )^{96} - 2 \left ( \frac{\sigma}{r} \right )^{48} \right ],
\end{align}
where $r$ the center-to-center distance, and $\epsilon$ the interaction strength is set to $20 k_{\mathrm{B}} T$. This is a smooth approximation of the well-known Asakura-Oosawa interaction~\cite{miyazaki2022asakura} in combination with steric repulsion. The beads comprising the interface (\textit{i.e.}, of our bead-spring bubble) can interact with the colloids forming the gel \textit{via} the same potential $ V_{\mathrm{LJ}}^{\mathrm{he}}$ with one modification. The interaction strength of the bead-colloid potential is appropriately rescaled to reproduce the effective depletion interaction between a colloid and the bubble surface (roughly twice that present between the colloids themselves; using a flat-wall approximation).

Our simulations were performed in periodic, cubic boxes with an edge length $L \in [50 \sigma, 120 \sigma]$. In each simulation, we used a volume fraction $\phi = 0.44$. This is a rather high value for colloidal gelation, but was chosen to closely approximate that of the experiment~\cite{Garbin-gel}. The bubble radius at rest was chosen to be $R_0 \equiv \langle R \rangle \in [10 \sigma, 40 \sigma]$. This choice departs from the value of the experiment --- the ratio of colloid-to-bubble radius therein is $\approx 10^2$ --- but proved necessary to achieve a desired computational efficiency. In experiment~\cite{Garbin-gel}, the curvature of the bubble is therefore lower than in our simulations, and the colloids near the interface therefore interact with an almost flat surface. We will return to the consequences of this choice in our discussion.

The gel was prepared \textit{via} an instantaneous deep quench from a purely repulsive potential to one with the aforementioned $20 k_{\mathrm{B}}T$ attraction strength. We allowed the gel to form for $50 \tau_{\mathrm{B}}$, where $\tau_{\mathrm{B}} = \sigma^2/(4D)$ is the Brownian time of the colloids with single-particle translational diffusion coefficient $D$. During this time, the bubble was left unperturbed, in order to allow the system form the gel network and relax internal stresses. Figure~\ref{fig:instance}a shows a representative snapshot of the initial configuration. After preparation of the bubble-gel system, the bubbles, were made to oscillate for 50 cycles, with different values of the frequency $\omega \in [2\pi/\tau_{\mathrm{B}}, \ 10^{5}2\pi/\tau_{\mathrm{B}}]$ and the oscillation amplitude ${\Delta R} \in [\sigma, \ 5\sigma]$. The oscillations were induced applying a sinusoidal perturbation on top of the equilibrium bubble pressure $p(t)=p_0(1+\delta \sin{\omega t})$. All simulations were performed using HOOMD-blue, a GPU-compatible Python package developed in the Glotzer Lab~\cite{ANDERSON2020109363}. 

\section{\label{sec:char}Characterization}

\begin{figure}[!htb]
 \centering
 \includegraphics[width=85mm]{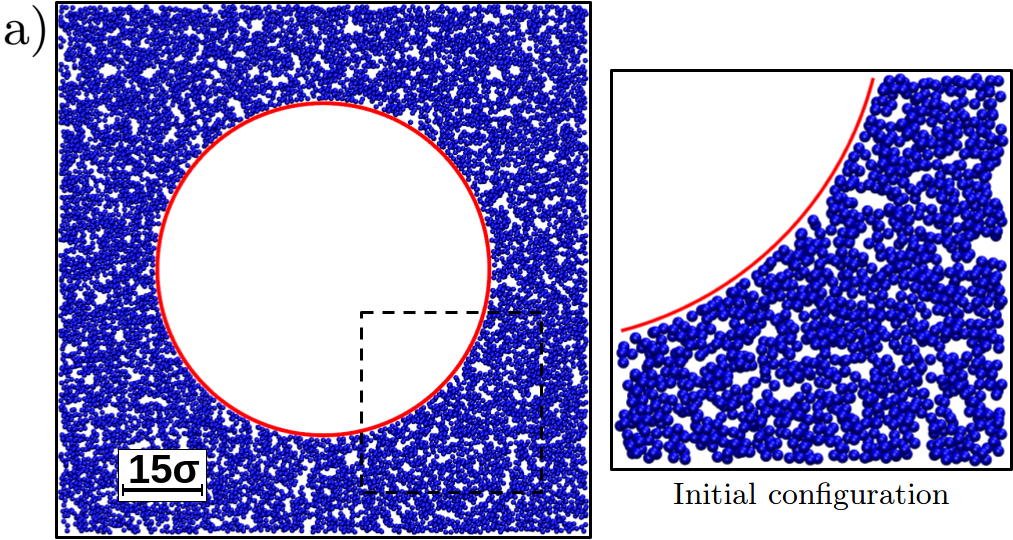} \\~\\
  \includegraphics[width=85mm]{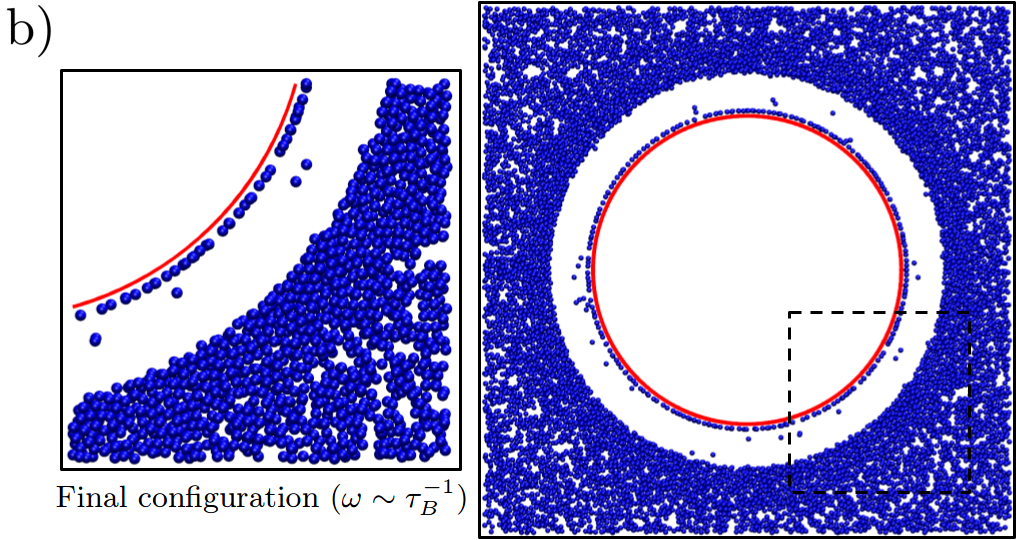} \\~\\
 \includegraphics[width=85mm]{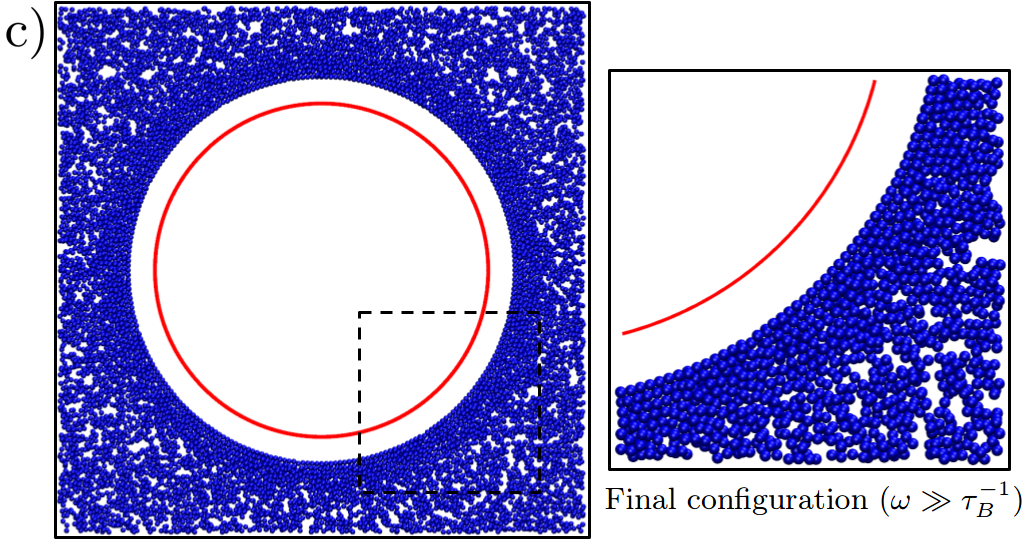}
 \caption{\label{fig:instance}The effect of bubble oscillations on the surrounding colloidal gel.  The bubble radius at rest $R_{0} \approx 30\sigma$ and the oscillation amplitude $\Delta R \approx 4\sigma$. Each panel shows a representative snapshot of the system. The main element is a vertical slice through the system that avoids the defects in the geodesic sphere (red circle) and has a width of $2\sigma$. This is complemented by a zoom-in (dashed square) on the region where the colloidal gel (blue) is most strongly distorted. Panel (a) shows the starting configuration, while (b) and (c) show a steady-state configurations (after $50 \tau_{\mathrm{B}}$) for values of frequency $\omega = 2 \pi / P$ with periods $P = 10^{-1} \tau_{\mathrm{B}}$ and $P = 10^{-5} \tau_{\mathrm{B}}$ expressed in the Brownian time $\tau_{\mathrm{B}}$, respectively.}
\end{figure}

We observed that our model bubble's motion modified the structure of the surrounding gel as follows. Figure~\ref{fig:instance} shows a representative snapshot of the initial and steady-state configurations that we obtained for small and large angular (oscillation) frequencies $\omega$ compared to the inverse Brownian time; oscillation amplitude $\Delta R = 4 \sigma$. In both cases, a void was formed between the gel and the bubble (at rest), which in experiment would be filled with fluid. Further out from the bubble, the colloid density visibly increased. At the largest distances the gel network appeared unperturbed. For $\omega \sim \tau_{\mathrm{B}}^{-1}$ the denser region appears disordered (Fig.~\ref{fig:instance}b), while for $\omega \gg \tau_{\mathrm{B}}^{-1}$ (Fig.~\ref{fig:instance}c) the dense is clearly ordered. 

For the lowest applied frequencies, we even observed rupture of the gel network, as evidenced by a layer of colloids that had become attached to the bubble surface due to the depletion interaction. Also note that some of the colloids have become detached from the gel and are freely floating in the `fluid-filled' void between the bubble surface and the gel in Fig.~\ref{fig:instance}b. The results presented in Fig.~\ref{fig:instance} suggest a connection between frequency of oscillation and reordering in the colloidal gel. We quantified this using averaged local bond-order parameters~\cite{lechner2008accurate} (BOP). These are non-dimensional parameters that can be used to distinguish ordered structures from disordered ones. In particular, we choose $q_6$ as indicator of reordering in the system, as it is the most significantly affected by the bubble oscillations.

\begin{figure}[!htb]
 \centering
 \includegraphics[width=89mm]{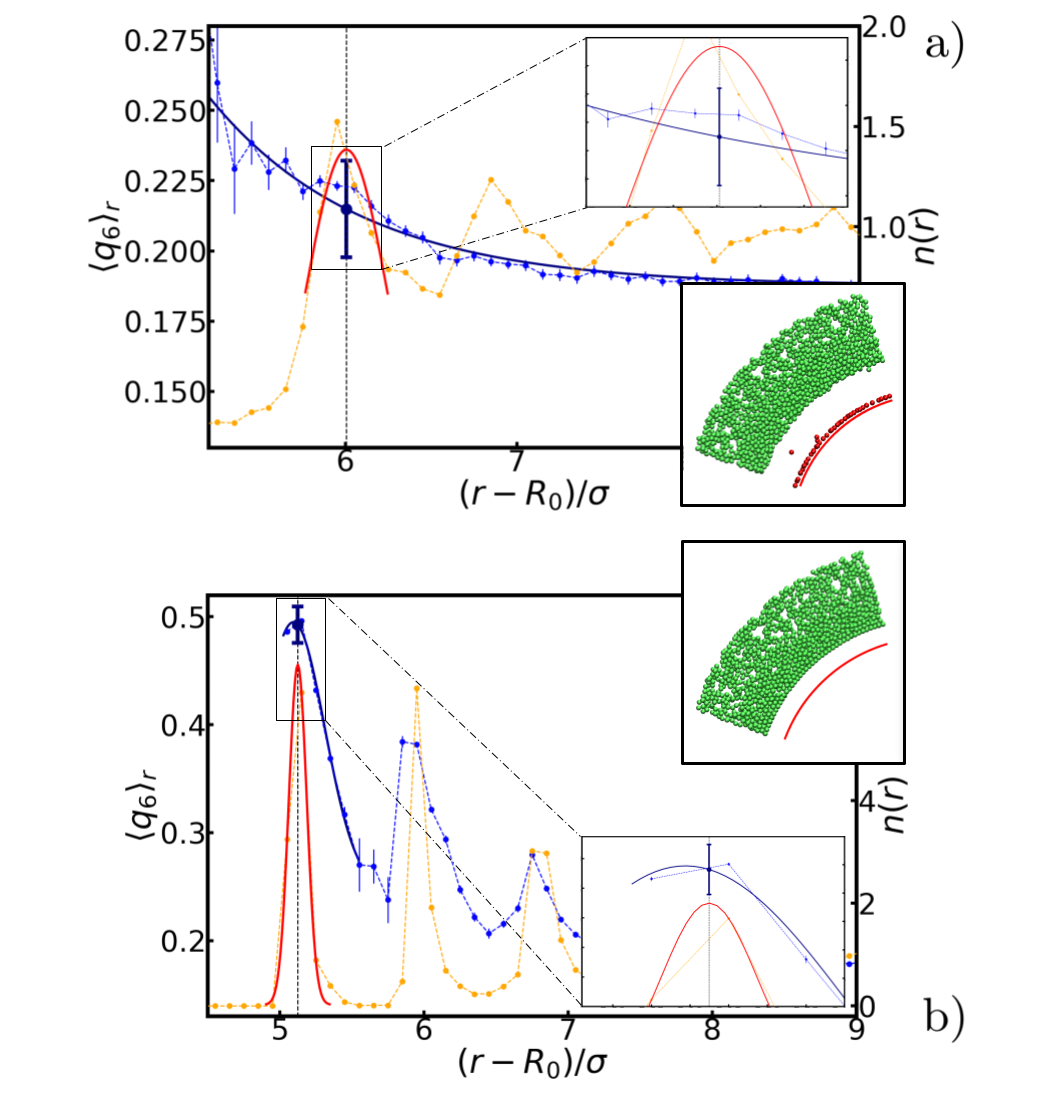}
 \caption{\label{fig:excludpart}Quantization of the ordering effect of the bubble oscillations on the surrounding gel. The radial density function $n(r)$ (yellow) and the radially averaged $q_{6}$ bond-order parameter $\langle q_{6} \rangle_{r}$ (blue). The symbols provide our data points and the error bars indicate the standard error of the mean. The blue and red fitted curves are used in our analysis procedure and described in the main text. Through the fit procedure, we locate the position of the first peak in $n(r)$, as indicated using the vertical gray dashed line. The second set of insets shows a wedge from a snapshot taken at steady state. Green particles remained part of the gel network, while red particles were either in a gas phase or has become attached to the bubble surface (represented here by the red arc). Detached particles are not considered in our analysis, as they do not contribute to the gel microstructure. Panels (a) and (b) show the steady-state profiles for a bubble-oscillation angular frequency of $\omega = 20 \pi / \tau_{b}$ and $\omega = 10^{5} 2 \pi / \tau_{b}$, respectively. For both (a) and (b) the bubble radius at rest is $R_0 \approx 30\sigma$ and the oscillation amplitude is $\Delta R \approx 4\sigma$.}
\end{figure}

Given the symmetry of the system, we made use of a radial average $\langle q_6 \rangle_{r}$, \textit{i.e.}, we measured the quantity shell by shell. The effect was pronounced around those layers that are (in temporary) contact with the bubble, and we therefore focused on the first few intact particle shells, as measured from the center. Two representative results are shown in Fig.~\ref{fig:excludpart}. We compared the values before and after the oscillations, and choose to represent each configuration with a single $\langle q_{6} \rangle$ value taken at distance $r^{\ast}$ and denoted henceforth by $\langle q_{6} \rangle_{r^{\ast}}$. Here, $r^{\ast}$ is the value, at which the radial density function $n(r)$ has its first peak \textit{and} the radially averaged coordination number $\langle z \rangle_{r^{\ast}} \ge 6$. The introduction of $\langle q_{6} \rangle_{r^{\ast}}$ will prove useful in constructing our state diagrams.

The appendix provides the details of our procedure to arrive at $\langle q_{6} \rangle_{r^{\ast}}$. In brief, we fitted each peak to $n(r)$ using a Gaussian function to determine $r^{\ast}$. For $\langle q_{6} \rangle_{r}$, we instead used a decaying exponential for disordered configurations (\textit{e.g.}, Fig.~\ref{fig:instance}b), and a Gaussian function for ordered ones (\textit{e.g.}, Fig.~\ref{fig:instance}c). We computed the standard error of the mean by summing uncertainties in the data and variances from the fitted functions, with the former being typically negligible compared to the latter.

Lastly, we quantified the length scale associated with the reordering in the system by fitting the peaks in $\langle q_{6} \rangle_{r}$ (when present) together with the values in the bulk. We found that the extent of the restructuring is roughly double the amplitude of oscillations,~\textit{i.e.}, $2 \Delta R$. The derivation of this typical length and the detail of the fits are provided in the appendix.

\section{\label{sec:res}State Diagram}

\begin{figure}[!htb]
 \centering
 \hspace{6mm} 
 \includegraphics[width=79mm]{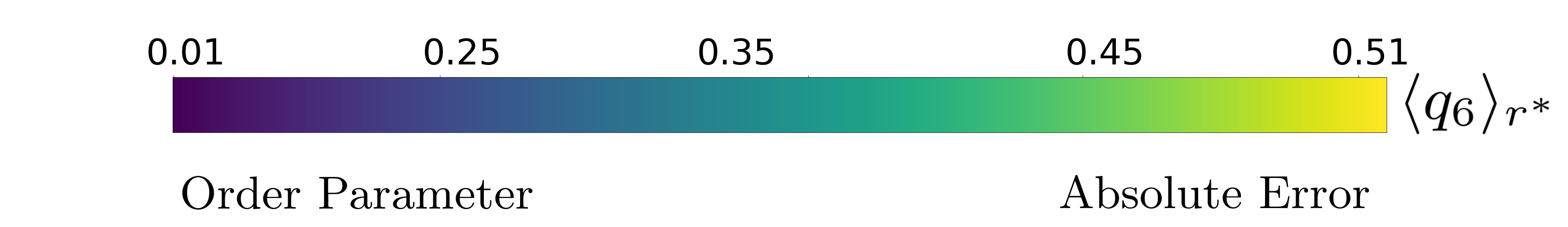}
 \includegraphics[width=44.75mm]{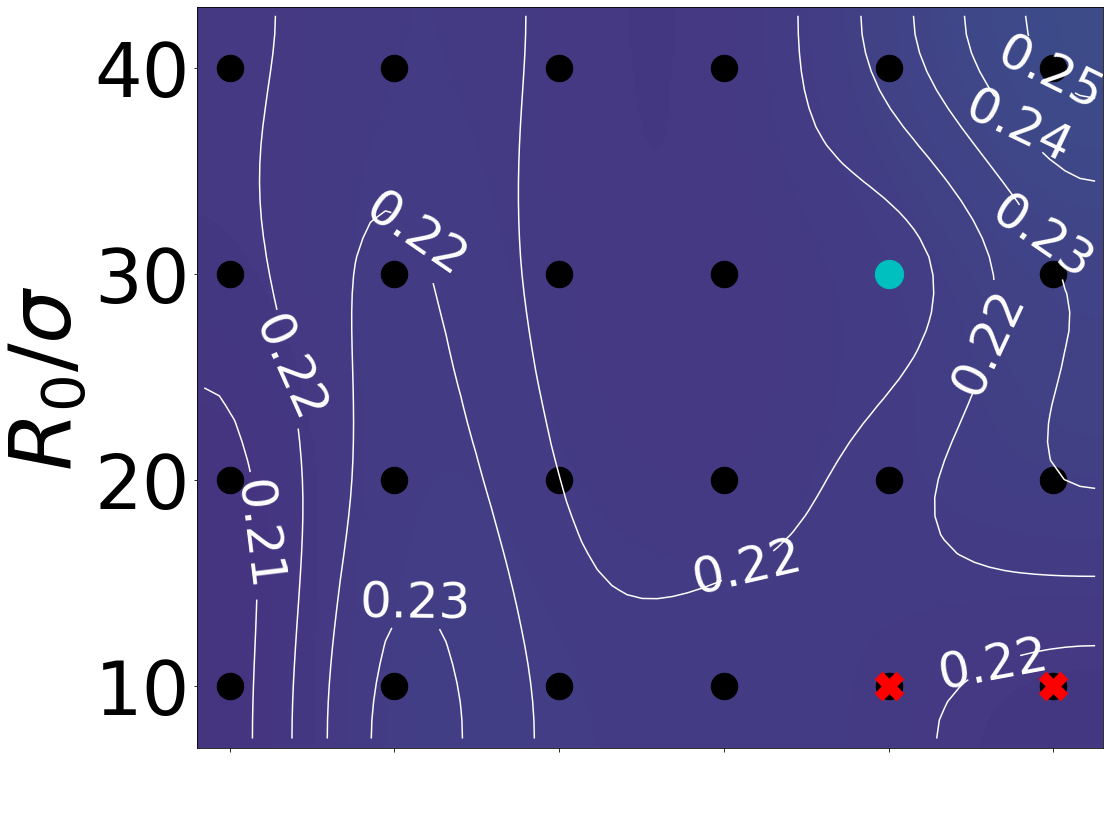}
 \includegraphics[width=40mm]{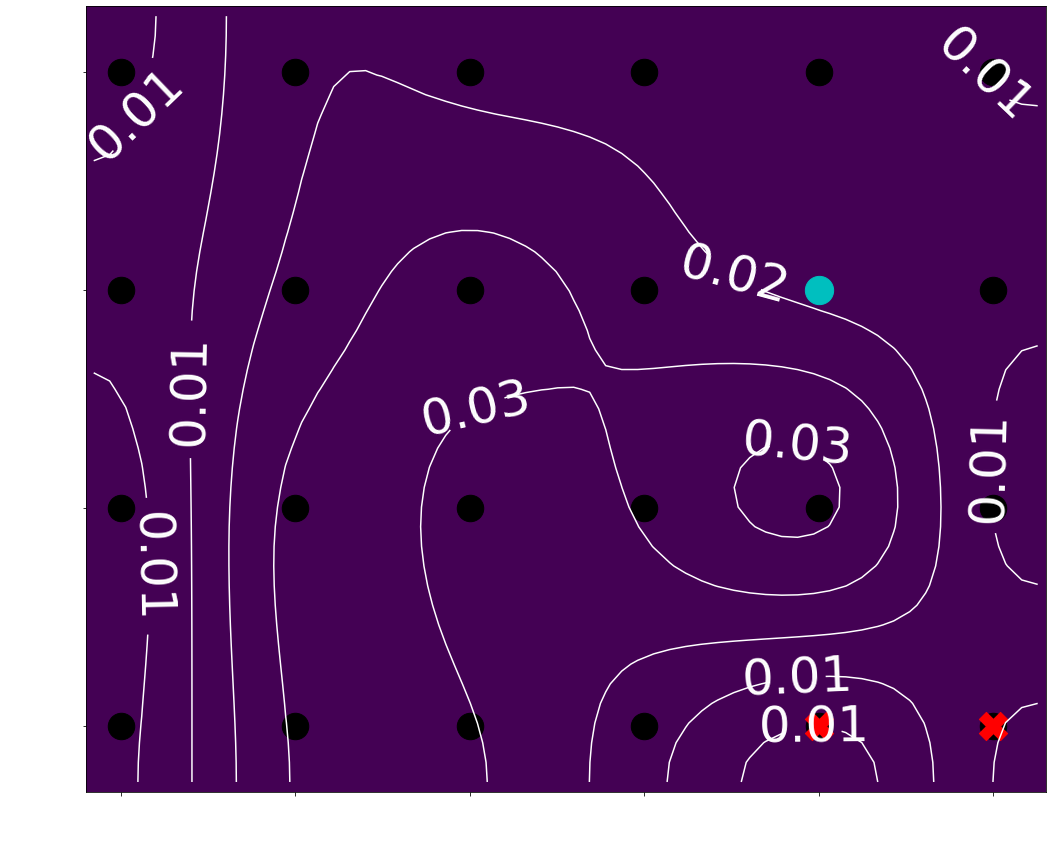}
 \includegraphics[width=44.75mm]{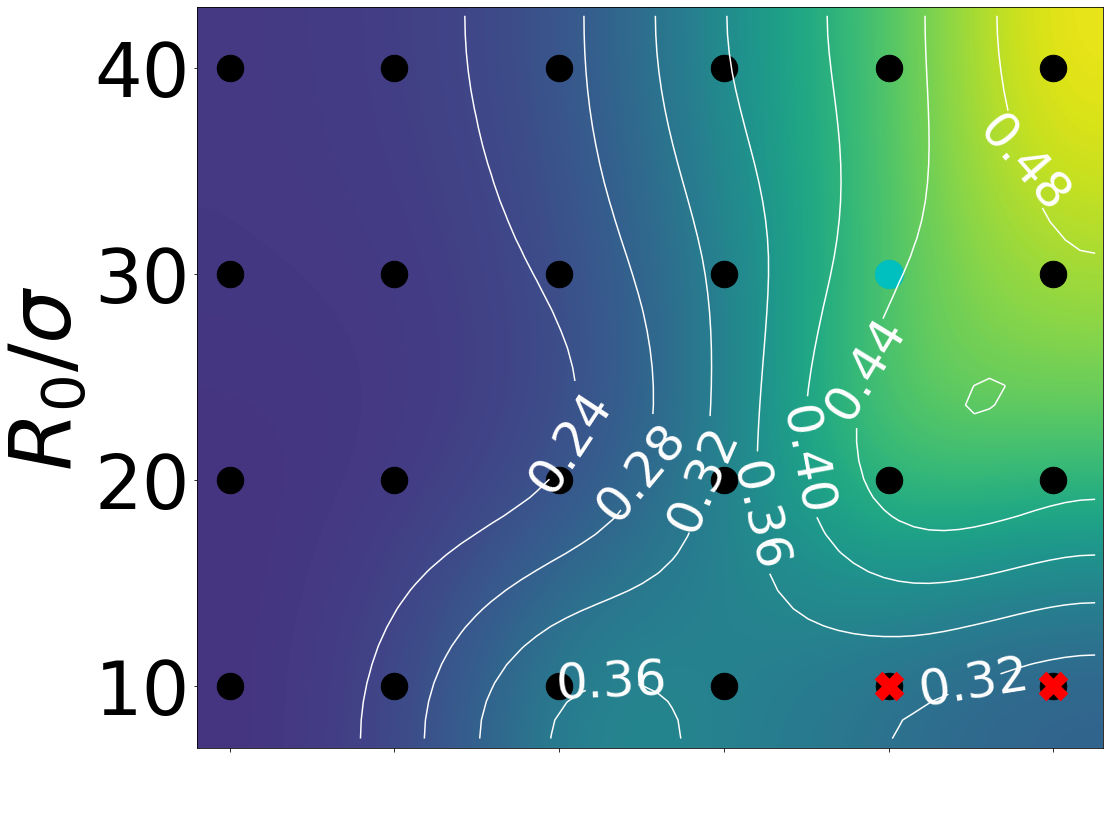}
 \includegraphics[width=40mm]{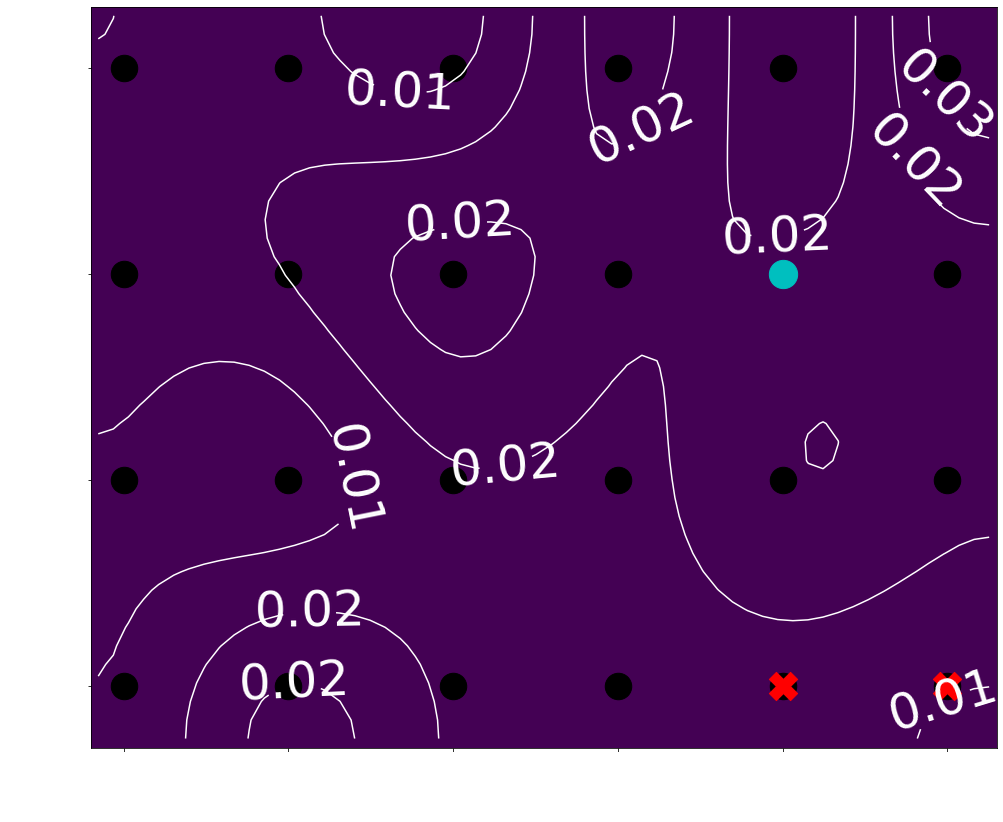}
 \includegraphics[width=44.75mm]{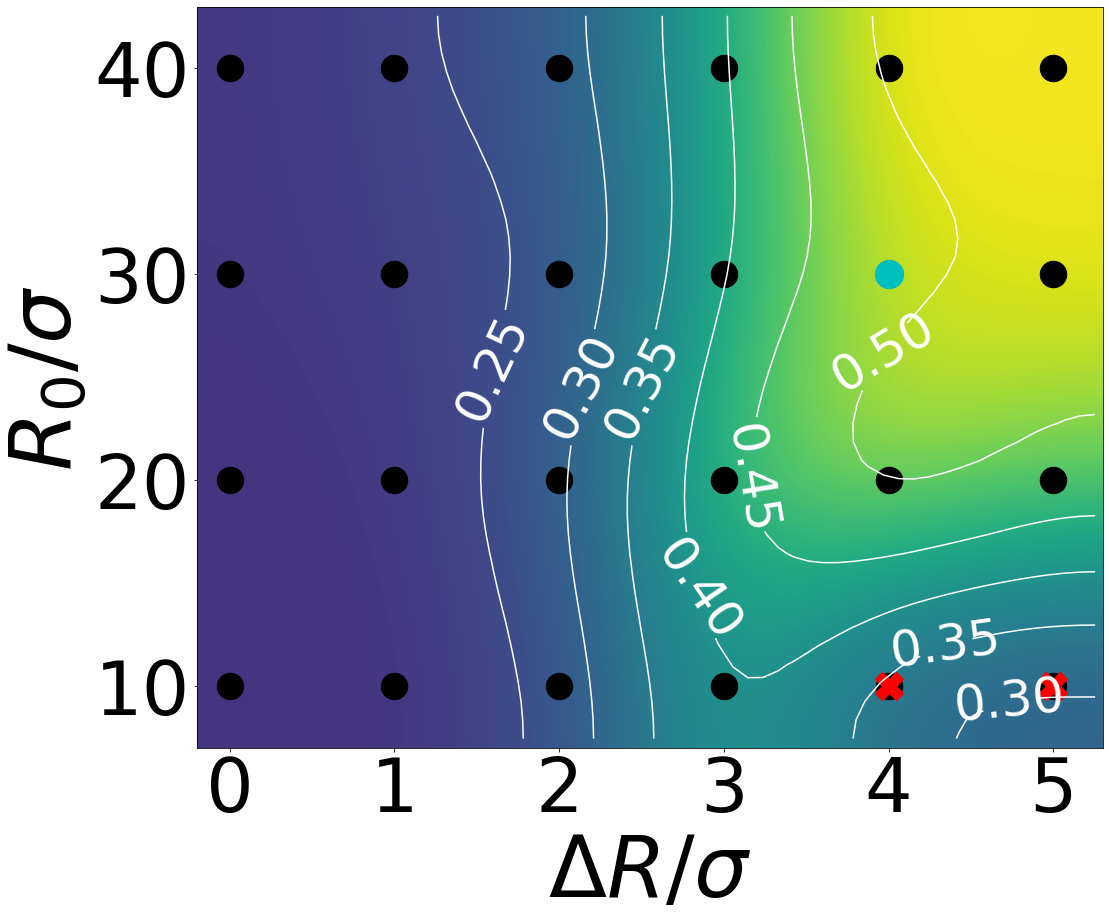}
 \includegraphics[width=40mm]{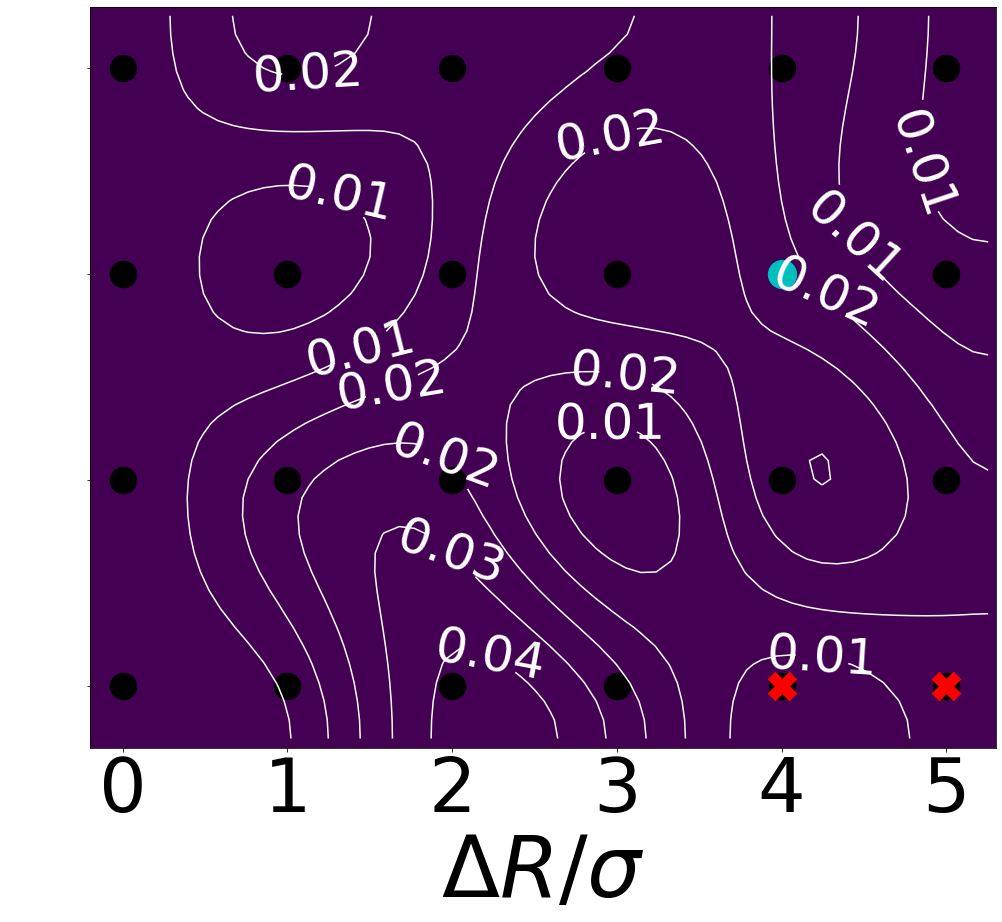}
 \caption{\label{fig:phasediag}Charting the effect of bubble radius $R_{0}$ and oscillation amplitude $\Delta R$ on the restructuring of the gel. The panels come in pairs (a) and (b), \textit{etc.}, and respectively show (left) the value of 6-fold, bond-order parameter in the first shell $\langle q_{6} \rangle_{r^*}$ and (right) the absolute error therein. From top to bottom the angular frequencies of the bubble oscillations are $\omega=2\pi / 10^{-1} \tau_{\mathrm{B}}$, $\omega=2\pi / 10^{-3}  \tau_{\mathrm{B}}$, and $\omega=2\pi / 10^{-5} \tau_{\mathrm{B}}$. The points where we obtained our data, are indicated using (black) dots. A smooth interpolation was created using a bicubic scheme (color map), for which the thin white curves provide iso-value contours. Data points marked with a red cross are excluded from our analysis because they belong to regions that are biased by the bubble tessellation, as explained in the main text. The data points highlighted in cyan represent configurations also analysed in Fig~\ref{fig:phasediagw}.}
\end{figure}

The above quantitative analysis allowed us to map the explored configurations onto a state diagram. Figure~\ref{fig:phasediag} shows three such mappings, providing the enhancement in local order using the steady-state $\langle q_{6} \rangle_{r^{\ast}}$ as a function of the bubble radius $R_{0}$ and oscillation amplitude $\Delta R$ (both normalized by the colloid diameter $\sigma$). We show the result for a low, an intermediate, and a large $\omega$ compared to $\tau_{\mathrm{B}}^{-1}$. We confirm the absence of constructive (increasing $q_{6}$) restructuring for the low-frequency configurations $\omega=2\pi / 10^{-1} \tau_{\mathrm{B}}$. For intermediate values of frequency $\omega=2\pi / 10^{-3}  \tau_{\mathrm{B}}$, we see the emergence of a wide region in the phase diagram where $\langle q_{6} \rangle_{r^*}$ reaches values that are slightly over twice ($\approx$ 0.5) those of the initial configurations ($\approx$ 0.2). The highest frequency regime shows the same features, but with an even wider zone of restructuring in the state diagram. This further supports the idea that local crystallization of the gel is strongly controlled by the frequency of oscillation.

The data in Fig.~\ref{fig:phasediag} allows us to conclude that restructuring is triggered only for sufficiently large bubbles ($R_0 \gtrsim 20 \sigma$). Additionally, restructuring in the gel is only possible, when the bubble can sufficiently expand and contract the surrounding gel. As there are no prescribed long-ranged interactions in our simulations (no hydrodynamic flows), compression in the gel is entirely dictated by the oscillation amplitude. We find that ordered structures can only emerge for $\Delta R \gtrsim 3\sigma$.

Here, we should note that there are small areas in Fig.~\ref{fig:phasediag} corresponding to small $R_{0} / \sigma$, yet large $\Delta R / \sigma$, that appear not to be affected by the bubble motion. This is an artifact of our bubble model: the surface discretization becomes comparable to the colloid-colloid separation. This gives rise to an effective egg-carton-like potential (for large values of $\Delta R/R_{0}$) that induces preferential colloid positions at the interface, which would not be present for a molecular interface. This effective bubble-colloid interaction interferes with structuring.

Avoiding these artifacts, we realize that $R_{0}/\sigma$ ratio determines the geometry of the collision between the gel network and the expanding bubble. For $R_{0} \gg \sigma$, the network experiences an interaction with an almost flat surface. This favours the alignment of the colloids in the gel, promoting the formation of ordered structures.

\section{\label{sec:osc}Oscillation Frequency}

\begin{figure}[!htb]
 \centering
 \includegraphics[width=85mm]{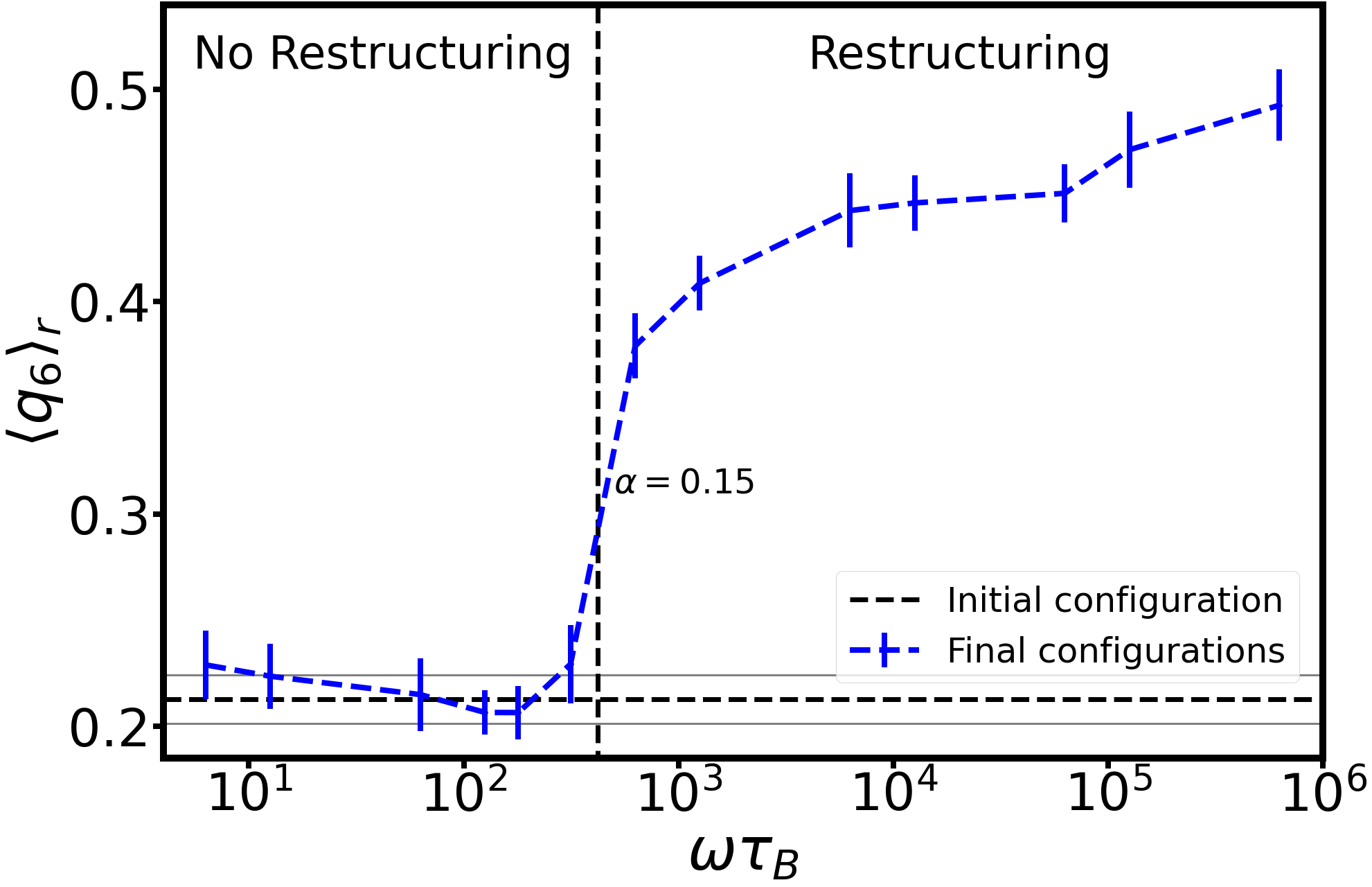}
 \caption{\label{fig:phasediagw}The effect of angular frequency $\omega$ on the restructuring of the gel around an oscillating bubble. Here, we choose $R_0 = 30\sigma$ and set the oscillation amplitude at $\Delta R = 4\sigma$. The peak value of the bond-order parameter $\langle q_{6} \rangle_{r^{\ast}}$ is given as a function of $\omega$ normalized by the Brownian time $\tau_{\mathrm{B}}$. The blue circles show the data points and the error bar provides the standard error of the mean. The horizontal dotted line indicates the $\langle q_{6} \rangle_{r^{\ast}}$ value of the initial configuration (before the bubble oscillations) and the two solid gray lines indicated the associated error. The dashed vertical line represents the frequency for which we localize a crossover between two regimes; $\alpha$ = 0.15, see main text.}
\end{figure}

We will focus on the large $R$, high $\omega$ situation next, as it more closely aligns with the experimental setup of Ref.~\cite{Garbin-gel}. By fixing $R_{0} = 30 \sigma$ and $\Delta R = 4 \sigma$, we can isolate the effect of the angular frequency $\omega$, see Fig.~\ref{fig:phasediagw}. We observe two trends separated by a relatively sharp transition in the value of $\langle q_{6} \rangle_{r^{\ast}}$ as a function of $\omega$. For small values of $\omega$ there was no restructuring, while for sufficiently large $\omega$ the value of $\langle q_{6} \rangle_{r^{\ast}}$ saturated to its crystalline result. This aligns with our analysis in Fig.~\ref{fig:phasediag}. The figure suggests that reordering in the gel is possible only if the time scale associated with bubble motion is negligible compared to thermal diffusion of the colloids.

To understand the role of $\omega$, we make an analogy to the frequency response of a colloidal gel under oscillatory shear~\cite{petekidis}. Using a Kramer's argument, the authors of Ref.~\cite{petekidis} estimated the effect of shear on the probability for a particle to escape from the attractive potential of its neighbor. Considering the typical escape time as a function of the shear frequency, they concluded that there is a critical threshold, below which the particles can rearrange (to form crystalline structures).

Motivated by this, we contrast the period of oscillation with two times scales in our system: the thermal diffusion time and a network-extraction time, respectively. The former is $\tau_{\mathrm{B}} \ge 2 \pi \omega^{-1}$ in all our simulations ($\tau_{\mathrm{B}} \omega \gg 1$ in Fig.~\ref{fig:phasediagw}) and we therefore deem it irrelevant. For the latter, we obtain the dimensionfree combination
\begin{align}
    \alpha = \frac{2 \pi \omega^{-1}}{t_{\mathrm{esc}}} = \frac{4 \pi \epsilon}{\gamma \sigma \omega} ,
\end{align}
where we estimated the escape time $t_{\mathrm{esc}} = \gamma \sigma/(2\epsilon)$. This follows from introducing reduced variables ($\boldsymbol{r} \rightarrow \sigma \Tilde{\boldsymbol{r}}$, $V_{\mathrm{LJ}}^{\mathrm{he}} \rightarrow \epsilon \Tilde{V}_{\mathrm{LJ}}^{\mathrm{he}}$) in the equation of motion for a single colloid interacting with the bubble, see Eq.~\eqref{eom}. Grouping the terms then gives rise to a natural time scale associated with the interaction potential

We first turn to the regime, in which there is no constructive reordering ($\alpha \ge 1)$. Here, the bubble shrinks slowly enough to allow the extraction of colloids at each cycle of the oscillation: clusters in the gel phase are continuously ripped apart and reformed and crystalline layers therefore cannot readily form. For the lowest frequencies applied in our simulations, we even observed the formation of a monolayer of colloids attached to the bubble surface (here through depletion).

For moderate and high frequencies ($\alpha \in [10^{-4},10^{-1}]$), the bubble moved too fast to extract particles from the surrounding gel. As a result, the colloids experience only a radially outward compression force, which slowly expands the surrounding gel medium. This allows for the reordering of the gel into a (locally) crystalline structure. We conclude that $\alpha$ is a meaningful parameter and that structuring is predominantly controlled by the bubble's inability to extract colloids from the network.

\section{\label{sec:disc}Discussion}

Our simulation results suggest that local tuning of the microstructure can be triggered by oscillations of deformable inclusions. However, there is a qualitative mismatch between our results and the ones obtained in the experiments of Ref.~\cite{Garbin-gel}. In the experiment, even small oscillation amplitudes are sufficient to rearrange many layers of colloids --- reordering was observed to cover an area approximately one bubble radius outward from the edge of the bubble.

It is possible that the missing ingredient in our simulations is long-ranged hydrodynamic interactions. The porous-medium flow produced by the oscillating bubble, has the potential to influence particles that are far away from the surface, provided a sufficiently large shear-P{\'e}clet number can be achieved~\cite{petekidis}. This would allow crystal structures to emerge at distances greater than we predict here. However, only accurate hydrodynamic simulation can elucidate to what extent local vibrations can induce this level of rearrangement.

The main difficulty in performing such simulations is the presence of a (moving) gas-liquid interface. Accounting for complex interfaces with large differences in viscosity is a challenge in computational fluid dynamics~\cite{numerical_artif_interface}. An approximate method to account for the interface would be to ignore the density and viscosity differences and use the presence of the tesselation points on the bubble surface to induce flows. This can, for instance, be done in the Rotne-Prager-Yamakawa (RPY) formalism \textit{via} the HOOMD-blue plugin developed by Fiore~\textit{et al.}~\cite{fiore2017rapid}. Such an approximation relies on the idea that the flows internal to the gel, rather than the presence of a gas-liquid interface control the physics of the rearrangement. The downside of this route is, however, that this is an uncontrolled approximation to the full hydrodynamic problem. That is, there is no means by which to readily refine it through the addition of higher-order terms.

\section{\label{sec:concl}Conclusions and Outlook}

Summarizing, using Brownian Dynamics simulations and a minimal model based on depletion interactions, we have quantified how an oscillating microbubble embedded in an attractive colloidal gel locally modifies the structure of the gel around its surface. The effect of the bubble dynamics can be constructive --- meaning that the gel locally crystallizes --- if the oscillation amplitude and the colloid-bubble size ratio are sufficiently large. The former controls the amount of compression exerted on the gel, and the latter determines the geometry of colloid-bubble collisions. Reordering is observed only in configurations where multiple layers of colloids are compressed, and where the colloids in the gel interact with an almost flat bubble surface.

We found the frequency of oscillation to be a control parameter in the restructuring. The bubble dynamics compete with both thermal and potential energies in the system. The main factor determining the formation of crystalline layers is the competition between time scales associated with the breaking of clusters of colloids and period of oscillation. Larger frequencies prevent the extraction of colloids from the gel network (a destructive effect of bubble oscillations) and allow the formation of ordered structures \textit{via} slow compression of the surrounding gel. This reordering typically extended into the bulk of the gel for a range equal to the oscillation amplitude.

The present work lays a solid foundation for understanding the impact of bubble oscillations on gel microstructure. This includes follow-up studies aimed at explaining recent experimental work in this direction.

\section*{Acknowledgements}
The authors acknowledge NWO for funding through OCENW.KLEIN.354. We are grateful to Dr.~Valeria Garbin for useful discussions and to Marjolein de Jager for input on the order-parameter analysis. Open data package containing the means to reproduce the results of the simulations available at: [DOI]

\bibliographystyle{aip}
\bibliography{reference}

\section{\label{sec:appen}Appendix}

In this appendix, we provide the details for various fitting procedures that we employed. We start by describing the procedure used to fit the radial density function $n(r)$, see Fig.~\ref{fig:excludpart}. For each configuration studied, we use a Gaussian function to fit the data points in the vicinity of the first peak and estimate the peak position $r^{\ast}$:
\begin{align}
    n(r)=a e^{-(r-r^{\ast})^2/2b^2}+n_{0},
\end{align}
with $a$, $b$, and $n_{0}$ fitting parameters that are not relevant to the analysis. If the first peak corresponds to a radial shell with an average coordination number smaller than 6, the peak is ignored, as it corresponds to either colloids attached to the bubble or in the gas phase (colloids floating in the `fluid-filled' void between the bubble surface and the gel). In those cases, the second peak is considered; the analysis is otherwise unaffected.

Turning to the peak value of the 6-fold bond-order parameter, $\langle q_{6} \rangle_{r^{\ast}}$, we include only data points in the vicinity of $r^{\ast}$ in our fit of $\langle q_{6} \rangle_{r}$. We use either an exponential function, if the configuration considered does not show reordering (see Fig.~\ref{fig:excludpart}a), or again a Gaussian function, in the case that crystalline structures are present in the system (see Fig.~\ref{fig:excludpart}b): 
\begin{align}
    \langle q_{6} \rangle_{r} = \begin{cases}
a_{1} e^{-(r-c_{1})^2/2 b_{1}^2}+d  &\text{ordered state} \\
a_{2} e^{-b_{2} r}+c_{2} &\text{disordered state}
\end{cases} ,
\end{align}
where the constants $a_{i}$, $b_{i}$, $c_{i}$, and $d$ are fitting parameters. Subsequently, we evaluate $\langle q_{6} \rangle_{r}$ for $r=r^{\ast}$, thereby obtaining a value that we deem representative of the amount of restructuring present in the dense layer.

\begin{figure}[!tb]
 \centering
 \includegraphics[width=85mm]{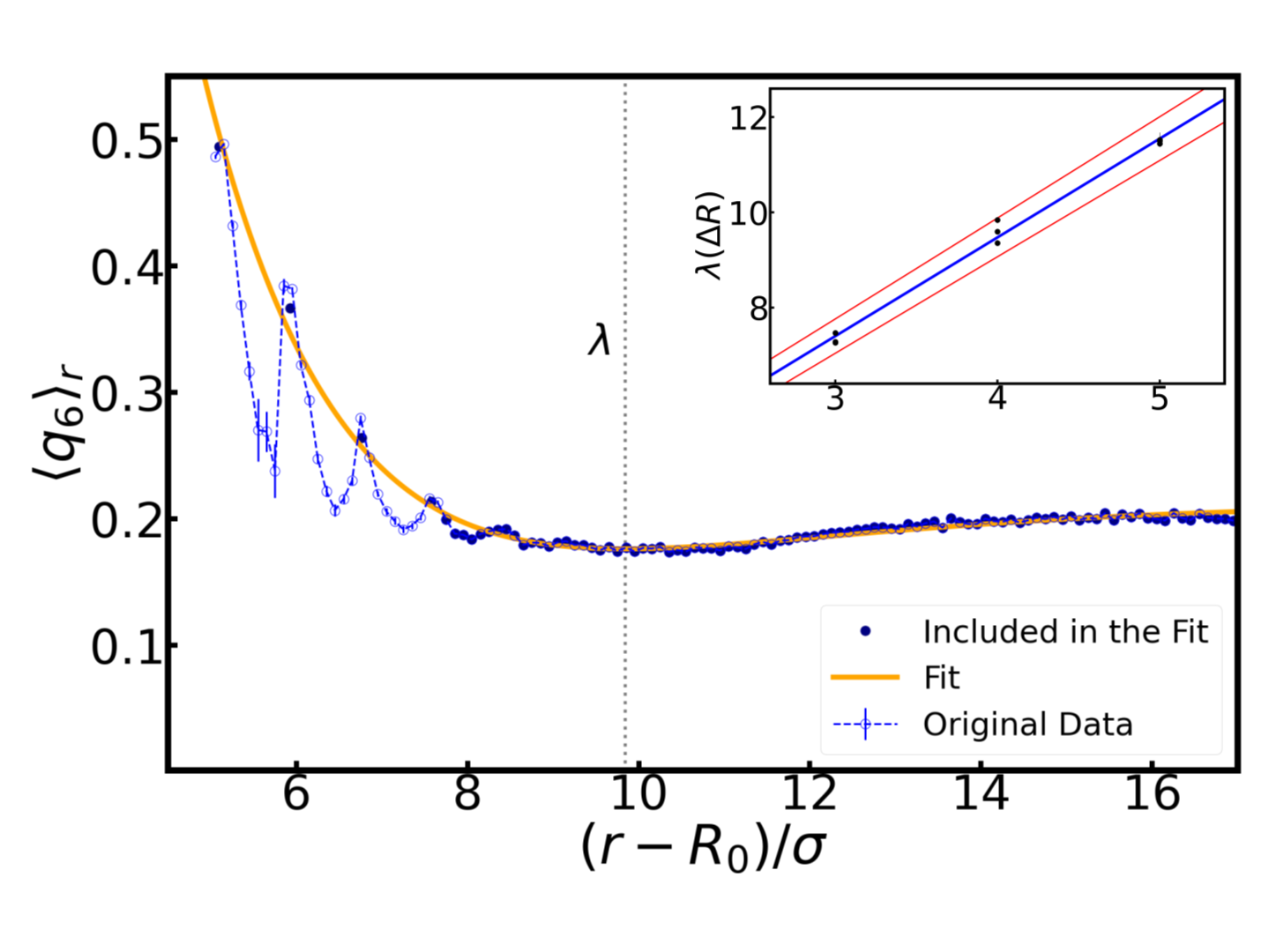}
 \caption{\label{fig:lambda}Characteristic length $\lambda$ over which the bubble oscillations reorder the surrounding gel. The original data set for $\langle q_{6} \rangle_{r}$ (blue circles); $R_{0}=30\sigma$, $\Delta R=4\sigma$, and $\omega=10^{5} 2 \pi / \tau_{b}$. The fitted function (orange curve) is computed using selected data from the original data set (highlighted in dark blue): the first $\Delta R$ peaks, obtained using Gaussian fits, as well as the bulk value of $\langle q_{6} \rangle_{r}$. The vertical dashed line indicates the position of the minimum of the fit, at which we place $\lambda$. The inset shows the linear trend --- the central value in blue, standard error of the mean delimited by the red lines --- obtained by all the computed values of $\lambda$ (9 in total).}
\end{figure}

We obtained an estimate of the length scale over which the bubble oscillations modify the gel structure, by fitting $\langle q_{6} \rangle_{r}$ for each ordered configuration ($R_0 \ge 20 \sigma$, $\Delta R \ge 3\sigma$), combining linear and exponential functions:
\begin{align}
    \langle q_{6} \rangle_{r} =& \ a(r-\kappa)e^{-\xi r}+d.
\end{align}
From the fit parameters, we can extract a characteristic length $\lambda = 1/\xi + \kappa $ corresponding to the distance where $\langle q_{6} \rangle_{r}$ has its minimum, see Fig.~\ref{fig:lambda}. In this figure, we focus on configurations with the highest value of the frequency of oscillation ($\omega= 2 \pi 10^{5}  / \tau_{b}$), as these show the strongest crystallization effects.

The inset shows the dependency of $\lambda$ on $\Delta R$. A linear fit of the obtained values provides us with $\lambda(\Delta R) = \beta \Delta R + \nu$, where $\nu = 1.2 \pm 0.3$ and $\beta=2.07 \pm 0.07$. Here, $\beta$ indicates that when $\Delta R$ layers of colloids are crystallized, the next $\Delta R$ layers in the gel have reduced ordering, before saturating to the bulk structure. The value of $\nu$ indicates that reordering of a single radius of colloids is to be expected, even without oscillations.

\end{document}